\documentclass{aa} 
\usepackage{graphicx}
\begin{document}
   \title{A Stellar Flare during the Transit of the Extrasolar Planet OGLE-TR-10b (Research Note)}

   \subtitle{}

   \author{Bentley, S. J.\inst{1}, Hellier, C.\inst{1}, Maxted, P. F. L.\inst{1}, Dhillon, V. S.\inst{3},  Marsh, T. R.\inst{2}, Copperwheat, C. M.\inst{2}, Littlefair, S. P.\inst{3}}

   \institute{Astrophysics Group, Keele University, Staffordshire, ST5 5BG, U.K. \and
Department of Physics, University of Warwick, Coventry, CV4 7AL, U.K. \and
Department of Physics and Astronomy, University of Sheffield, S3 7RH, U.K.}

   \date{Accepted 9 August 2009}

   \abstract{We report a stellar flare occurring during a transit of the exoplanet OGLE-TR-10b, an event not previously reported in the literature. This reduces the observed transit depth, particularly in the \textit{u'}-band, but flaring could also be significant in other bands and could lead to incorrect planetary parameters.  We suggest that OGLE-TR-10a is an active planet-hosting star and has an unusually high X-ray luminosity.$^{1}$
     
}
\titlerunning{A Stellar Flare on OGLE-TR-10a}
\authorrunning{S. Bentley et al.}

   \maketitle


\section{Introduction}
The radii of transiting extrasolar planets are determined from the transit depth.  However, the transit lightcurve can be affected by star spots (Rabus et al. \cite{rabus09}) or by intrinsic stellar variability (Boisse et al. \cite{boisse}).  There have been five different published transit depths of OGLE-TR-10, which has resulted in some uncertainty in the value of the radius of OGLE-TR-10b. Transit depths range from $\sim0.99\%$ (Holman et al. \cite{holman07}) to $\sim2.20\%$ (Udalski et al. \cite{udalski}).  This relatively large range of transit depths has been attributed to photometric errors associated with overcrowding in the OGLE-TR-10 field (Pont et al. \cite{pont07}).  Uncertainties in the spectral model of the host star have also lead to a range of stellar, and hence planetary radius, estimates (Ammler von Eiff et al. \cite{newteff}).

In this paper we report a flare during a transit of OGLE-TR-10b, which affects the observed transit depth, particularly in the \textit{u'}-band.  We also discuss whether OGLE-TR-10a is an active star.

\section{Observations, Analysis and Results}
Observations were made with ULTRACAM (Dhillon et al. \cite{ultracam}) on the VLT UT3 (Melipal) telescope on June 10th 2007. ULTRACAM is a triple-beam CCD camera that obtains simultaneous images in SDSS \textit{u'},\textit{g'} and \textit{i'} filters; each detector consists of a $1024\times{}1024$ pixel, back-illuminated, thinned, Marconi 47-20 frame-transfer CCD, with a platescale of 0.15"$\rm{pixel}^{-1}$.  We acquired 2100 exposures over a period of 6 hours, with exposure times of 9.5 seconds per frame for the \textit{i'} and \textit{g'}-band frames and 70.6 seconds per frame for the \textit{u'}-band frames.  The observing conditions were good throughout.  The data were reduced using ISIS (Alard \cite{alard}), using the reduction strategies of Hartman et al. (2004).

\begin{figure}
 \centering
 \resizebox{\columnwidth}{!}{\includegraphics[angle=270]{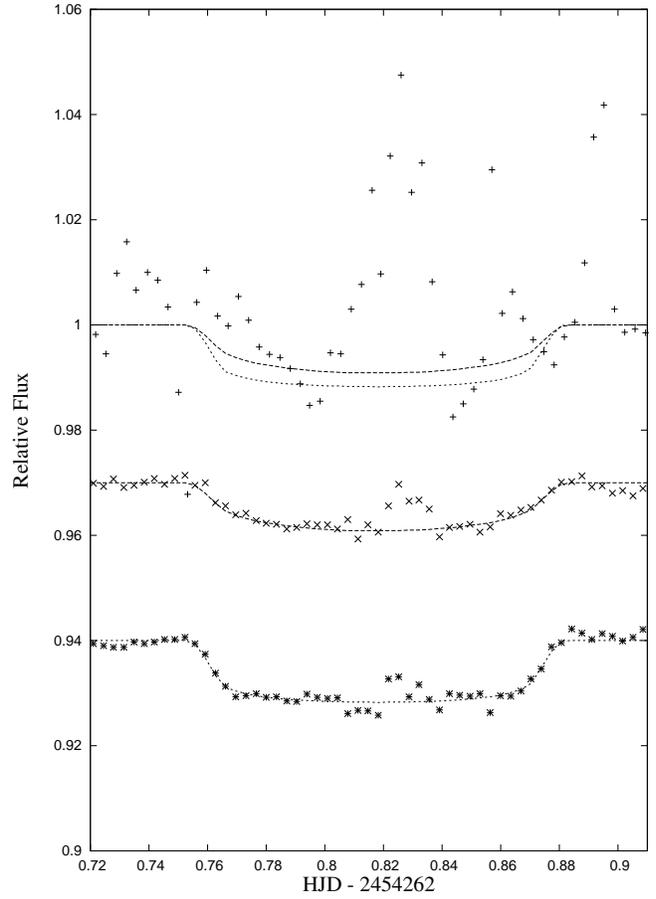}}
 \caption{\label{lcs} \small{Lightcurves, in 5 minute bins, of the transit of OGLE-TR-10 in the \textit{u'}-band (top), \textit{g'}-band (middle, offset by 0.03) and \textit{i'}-band (bottom, offset by 0.06).  The \textit{i'} and \textit{g'}-band lightcurves have the best-fitting transit overplotted; these fits are also overplotted in the \textit{u'}-band transit.}}

 \end{figure}

The lightcurves (Fig. \ref{lcs}) are contaminated by a flare during mid-transit between at least MJD=2454262.31 and 2454262.34.  The relative brightness of the flare increases from the \textit{i'}-band to the \textit{u'}-band, as expected for a small intensity flare (Tovmassian et al. \cite{tovmassian03}).

We used the Markhov-Chain Monte Carlo (MCMC) technique to estimate parameters of the system from the \textit{i'} and \textit{g'}-band lightcurves (we did not model the \textit{u'}-band lightcurve since the lightcurve is dominated by the flare).  We follow the general description given in Holman et al. (\cite{holman07}) using the analytical lightcurve modelling of Mandel \& Agol (\cite{mendelandagol}).  The free parameters in our fit are the radius ratio $R_{p}/R_{*}$, impact parameter $b$, time of central transit $T_{0}$, transit phase duration $R_{*}/a_{p}$ and $u_{1}$ and $u_{2}$, the non-linear limb-darkening coefficients ($R_{p}$, $R_{*}$ and $a_{p}$ are the planetary radius, stellar radius and semi-major planetary orbital axis, respectively).  We estimated our errors by using the scatter about the fit between the model and data, excluding the data points containing the flare.  We fixed the period to $P=3.10129$ days, as given by Pont et al. (\cite{pont07}).  To determine the effect of the flare on the transit depth, we computed fits both with and without the region of the lightcurve (HJD=2454262.81 to 2454262.84) containing the flare.

The results of our MCMC fit are shown in Table \ref{mcmc} and the lightcurves in Fig. \ref{lcs}.  In the \textit{i'}-band the $R_{p}/R_{*}$ is the same ($R_{p}/R_{*}=0.106^{+0.002}_{-0.003}$) regardless of whether the flare is excluded.  In the \textit{g'}-band the measured $R_{p}/R_{*}$ is slightly greater when excluding the flare region ($R_{p}/R_{*}=0.099^{+0.008}_{-0.006}$), than when including it ($R_{p}/R_{*}=0.093^{+0.005}_{-0.009}$).

\begin{table}
\begin{tabular}{lll}
	Parameter & \textit{i'} band value & \textit{g'} band value \\ \hline\hline
	\textit{$T_{0}$} (HJD) & $2454262.8175^{+0.0003}_{-0.0003}$ & $2454262.8180^{+0.0040}_{-0.0013}$ \\
	\textit{$R_{p}/R_{*}$} & $0.106^{+0.002}_{-0.003}$ & $0.093^{+0.005}_{-0.009}$ \\
	\textit{b} & $0.53\pm0.16$ & $0.83\pm0.15$ \\
	\textit{$R_{*}/a$} & $0.133^{+0.022}_{-0.017}$ & $0.185^{+0.058}_{-0.073}$ \\
	\textit{$u_{1}$} & $0.2491\pm0.0100$ & $0.6435\pm0.0050$ \\
	\textit{$u_{2}$} & $0.3204\pm0.0100$ & $0.1839\pm0.0050$ \\
\end{tabular} 
	\caption{\label{mcmc} MCMC-estimated parameters for the \textit{i'}-band and \textit{g'}-band lightcurves.}
\end{table}

\section{Discussion}
\subsection{The Radius Ratio}
There has been much disagreement in the literature over the reported $R_{p}/R_{*}$ of OGLE-TR-10; published $R_{p}/R_{*}$ values include the original OGLE measurement of 0.148 (Udalski et al. \cite{udalski}), $0.129\pm0.003$ (Bouchy et al. \cite{bouchy}), $0.127\pm0.017$ (Konacki et al. \cite{konacki}) and $0.099\pm0.002$ (Holman et al.  \cite{holman07}).  Attempting to solve this discrepancy, Pont et al. (\cite{pont07}) concluded that systematic noise was affecting the $R_{p}/R_{*}$ measurement of Udalski et al. (\cite{udalski}) and Holman et al. (\cite{holman07}).  Reporting new VLT data, Pont et al. (\cite{pont07}) argued for a radius ratio of $R_{p}/R_{*}=0.110\pm0.002$.  There appears to be no obvious signs of flares in any of the aforementioned authors' lightcurves.

Our \textit{i'}-band $R_{p}/R_{*}$ is compatible with the value of Pont et al. (\cite{pont07}).  Our \textit{g'}-band $R_{p}/R_{*}$, however, is the shallowest yet reported for OGLE-TR-10, which could result from contamination by the flare.  Our \textit{u'}-band lightcurve shows a transit that is nearly entirely filled in by the flare.

The variation of $R_{p}/R_{*}$ in each band is too large to be due to wavelength-dependant absorption in the atmosphere of OGLE-TR-10b.  In the \textit{i'} bandpass there is expected to be strong molecular absorption in the atmosphere of OGLE-TR-10b, although this will change the relative transit depth by no more than $10^{-4}$ (Carter et al. \cite{carter}).

\subsection{The Activity of OGLE-TR-10a}
Stellar X-ray emission is associated with stellar activity and planet-harbouring stars are found to be more X-ray active than non-planet-harbouring stars by a factor of $\sim4$ (Kashyap et al. \cite{kashyap08}).  Kashyap et al. (\cite{kashyap08}) also found that OGLE-TR-10a is among the most X-ray luminous planet-hosting stars with an X-ray luminosity of $\log{L_{X}}=30.34\pm0.25$ ergs $\rm{s^{-1}}$ which is unusually high for a G-dwarf (for the Sun, $28>\log{L_{X}}>27$ ergs $\rm{s^{-1}}$).  Furthermore, OGLE-TR-10a has an $L_{X}$ similar to RS CVn stars, which supports the notion that OGLE-TR-10a is an active star (Holmberg et al. \cite{holmberg}).

The estimated brightness of the flare is $55.24\pm3.50$ mmag for the \textit{u'}-band (using the \textit{i'}-band transit depth as an out-of-flare flux reference point), $5.45\pm1.50$ mmag for the \textit{g'}-band and $2.34\pm2.00$ mmag for the \textit{i'}-band, which are low amplitudes compared to other small flare events reported in the literature (e.g. Ventura et al. \cite{ventura}).  By estimating the quiescent luminosity of OGLE-TR-10a and the increase in magnitude we used the relations discussed in Gershberg \& Shakhovskaya (\cite{gershberg83}) to estimate the energy of the flare to be $E\approx(1\pm0.1)\times10^{32}$ ergs $\rm{s^{-1}}$.  Compared to solar activity, this is an energetic event, equivalent to an X13 class solar flare.

\begin{acknowledgements}
S.J.B. acknowledges the support of a STFC studentship.  ULTRACAM is supported by STFC grant
PP/D002370/1.  Based on observations collected at ESO, Chile.  We also thank Rob Jeffries for discussions on stellar age and X-ray activity.
\end{acknowledgements}

\end{document}